\documentclass[fleqn,twoside]{article}
\usepackage{espcrc2}

\usepackage{graphicx}
\usepackage[figuresright]{rotating}

\newcommand{\AmS}{{\protect\the\textfont2
  A\kern-.1667em\lower.5ex\hbox{M}\kern-.125emS}}

\title{Spectrum of screening masses in the (3+1)D
SU(2) pure gauge theory near the critical temperature~\thanks{Talk presented by A. 
Papa.}}

\author{Roberto Fiore\address[CS]{Dipartimento di Fisica, Universit\`a della 
Calabria \\ \& Istituto Nazionale di Fisica Nucleare, Gruppo Collegato di Cosenza,
Italy},
Alessandro Papa\addressmark[CS]
and Paolo Provero\address{Dipartimento di Scienze e Tecnologie Avanzate, 
Universit\`a del Piemonte Orientale, Alessandria, \\
Dipartimento di Fisica Teorica, Universit\`a di Torino \\ 
\& Istituto Nazionale di Fisica Nucleare, Sezione di Torino, Italy}}
       
\begin{document}

\begin{abstract}
We study the spectrum of screening masses in the deconfined phase of
(3+1)D SU(2) pure gauge theory near criticality and compare it with the
spectrum of bound states in the broken symmetry phase of the 3D Ising
model, which is related to the gauge theory by universality.
\vspace{1pc}
\end{abstract}

\maketitle

\section{INTRODUCTION}
It is well known that both the 3D Ising model
\[
S=-\beta \sum_{<n,m>} s_n s_m
\]
and the 3D lattice regularized $\phi^4$ theory
\[
S=-\beta \sum_{<n,m>} \phi_n \phi_m + \sum_n \phi_n^2+\lambda \sum_n (\phi_n^2
-1)^2
\]
undergo a second order phase transition at $\beta=\beta_c$ from a 
phase where the Z(2) symmetry is broken ($\beta>\beta_c$) to a phase when 
the Z(2) symmetry is restored ($\beta<\beta_c$). The order parameter of 
this transition is the expectation value of the fundamental field, i.e. the 
magnetization.

The spectrum of massive excitations in the broken symmetry phase of 3D Ising model 
and $\phi^4$ theory can be determined from the analysis of
wall-wall correlations of spin operators. By performing Monte Carlo simulations 
near $\beta_c$, it has been possible to detect three $0^+$ states~\cite{CHP99}.
Results for their mass ratios agree in the two models: $m_2/m_1=1.83(3)$ and
$m_3/m_1=2.45(10)$ ($m_1$ is the fundamental mass). Since the 3D Ising model
and the 3D (lattice) $\phi^4$ theory belong to the same universality class, this
result shows that universality is valid far beyond critical indices.

The first excitation ($m_2$) is below the two particle production threshold,
it represents therefore a non-perturbative state, identified as a bound state
of the fundamental excitation~\cite{CHPZ01}. The second mass state ($m_3$) 
could be a fictitious one which mimics the two-particle cut.

According to {\em duality}, the broken phase of the 3D Ising model can be put in 
correspondence with the confined phase of the 3D Z(2) gauge theory. This implies
that mass ratios in the glueball spectrum in the confined phase of 3D Z(2) gauge 
theory should agree with mass ratios in the broken phase of 3D (spin) Ising model.
This has been verified in Ref.~\cite{CHPZ01} using Monte Carlo 
results~\cite{ACCH97} for the $0^+$ glueball masses in the 3D Z(2) gauge theory: 
$m_2/m_1=1.88(2)$ and $m_3/m_1=2.59(4)$.

According to {\em universality}, the broken phase of 3D Ising model corresponds to 
the deconfined phase of the (3+1)D SU(2) pure gauge theory at finite 
temperature~\cite{SY82}, with the 
magnetization of the spin model mapped into the Polyakov loop, $P(x,y,z)\equiv 
\mbox{Tr} \prod_{n_4=1}^{N_t} U_4(x,y,z, a n_4)$. This implies that the same mass
ratios observed in the broken phase of 3D Ising should appear in the spectrum 
of the slopes of the exponential decays of wall-wall correlations of 
Polyakov loop operators, i.e. in the spectrum of {\em screening masses},
in the deconfined phase of (3+1)D SU(2) at finite temperature near 
criticality. In this work we present a numerical test of this prediction.

\section{NUMERICAL RESULTS}
A naive way to determine $0^+$ screening masses is to study the connected 
wall-wall correlation function
\[
G(|z_1-z_2|) \equiv \langle \overline P(z_1) \overline P(z_2) \rangle 
- \langle \overline P(z_1) \rangle \langle \overline P(z_2) \rangle 
\]
\[
= A (e^{-m |z_1-z_2|} + e^{-m (N_z-|z_1-z_2|)}) \\
\]
\[
+ B (e^{-m^* |z_1-z_2|} + e^{-m^* (N_z-|z_1-z_2|)}) + \ldots \;.
\]
Here $\overline P(z)\equiv 1/(N_x N_y)\sum_{x,y} P(x,y,z)$ is the 
wall-average of Polyakov loops, $m$ is the lowest mass in the $0^+$ channel, 
$m^*$ the first excited level in the $0^+$ channel and the dots represent
the contribution from higher mass excitations and from multi-particle cuts.
Since $\langle \overline P(z) \rangle \neq 0$ in the high temperature 
phase, in principle also a $z$-independent term in $G(|z_1-z_2|)$ of the 
form $C_0 \exp[-m N_z]$ should be included (see~\cite{MW87}). However, such
term is evidently sub-dominant for $m^*<2m$, as we expect to be 
the case here.

\begin{figure}[htb]
\includegraphics[scale=0.37]{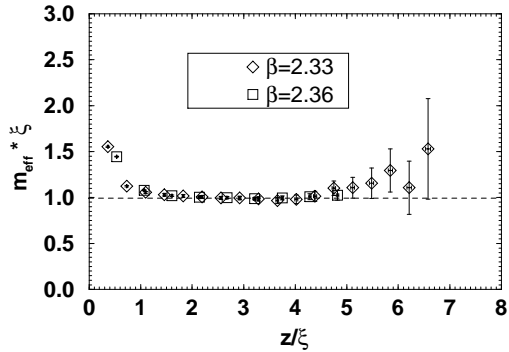}
\vspace{-0.8cm}
\caption{Effective screening masses in physical units at 
$\beta=2.33$ on a $18^2\times36\times4$ lattice and $\beta=2.36$
on a $18^3\times4$ lattice.}
\label{scaling}
\vspace{-0.7cm}
\end{figure}

We performed Monte Carlo simulations using the overheat-bath updating 
algorithm~\cite{PV91} with Kennedy-Pendleton improvement~\cite{KP85}.
We studied the naive wall-wall correlation at $\beta=2.33$ on a 
$18^2\times36\times4$ lattice (statistics 1M) and at $\beta=2.36$ on a 
$18^3\times4$ (statistics 1.5M). These values of $\beta$ are slightly 
above the critical value $\beta_c=2.29895(10)$ determined in Ref.~\cite{ES98}.  
We determined the effective mass defined as
\[
G(z) = C (e^{-m_{\mbox{\tiny eff}} z} + e^{-m_{\mbox{\tiny eff}} (N_z-z)})
\]
and found in both cases that data for $m_{\mbox{\tiny eff}}$ as a function of $z$
reach a plateau value corresponding to the fundamental mass: $m=0.3660(20)$ at 
$\beta=2.33$ and $m=0.5356(26)$ at $\beta=2.36$. The deviations from the plateau
at the smallest values of $z$ can be attributed to lattice artifacts and to the 
possible effect of other physical states. In order to single out the latter effect,
we rescaled the values of $m_{\mbox{\tiny eff}}$ and $z$ by $m\equiv1/\xi$ and 
$\xi$, respectively, and put together on the same plot data from $\beta=2.33$
and $\beta=2.36$. As shown in Fig.~\ref{scaling}, data from the two different 
$\beta$'s fall on the same curve {\em before} reaching the plateau, 
indicating the physical origin of these effects.

In order to detect this excited state, we adopted the so-called ``variational'' 
method~\cite{Kro90,LW90}. The method includes the following steps: define 
a basis of (wall-averaged) operators $\{O_\alpha\}$, compute the 
connected cross-correlation matrix among them
\[
C_{\alpha\beta}(z)=\langle O_\alpha(z)O_\beta(0)\rangle-
\langle O_\alpha(z)\rangle \langle O_\beta(0)\rangle\;,
\]
diagonalize $C_{\alpha\beta}(z)$ to obtain the eigenvalues
\[
\lambda_i(z)\propto e^{-m_i z}+e^{-m_i (N_z-z)}
\]
and finally extract the masses $m_i$. In practical simulations not more than two
(sometimes three) leading eigenvalues allow to extract a signal. The effectiveness 
of the method relies on the choice of a ``good'' set of operators; it is 
convenient to define operators living on different length scales
(for instance, by use of recursive definitions). Moreover, by defining 
operators with non-trivial transformation under spatial rotation, it is 
possible to look for states with non-zero angular momentum.

\begin{figure}[htb]
\vspace{-1cm}
\includegraphics[scale=0.37]{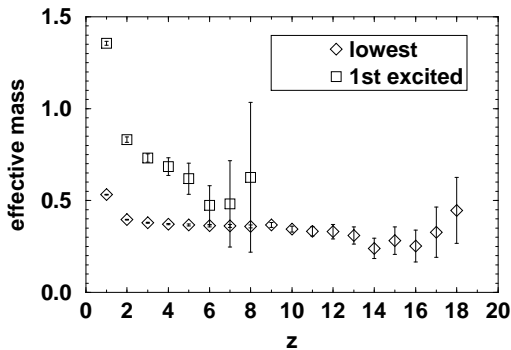}
\vspace{-0.9cm}
\caption{Lowest and first excited effective screening masses in the $0^+$ channel
at $\beta=2.33$ on a $18^2\times36\times4$ lattice from a set of six operators
(smoothing levels 0, 3, 6, 9, 12 and 15).}
\label{0+}
\vspace{-0.7cm}
\end{figure}

As a first set of operators we considered the one which can be built adapting
to the present case the recursive procedure used in Ref.~\cite{CHP99}: 
\[
P^{(0)}(x,y,z) = P(x,y,z)\;,
\]
\[
P^{(n+1)}(x,y,z) = \mbox{sign}(u) \biggl[(1-w)|u|+w y\biggr]\;,
\]
\begin{eqnarray*}
u&=&\frac{1}{4}\biggl(P^{(n)}(x-a,y,z)+P^{(n)}(x,y-a,z) \\
&+&P^{(n)}(x+a,y,z)+P^{(n)}(x,y+a,z)\biggr)\;,
\end{eqnarray*}
with $w=0.1$ and $y=\langle P(x,y,z) \rangle$. We considered a set of six
such (wall-averaged) operators corresponding to the smoothing steps
0, 3, 6, 9, 12 and 15. The effective masses extracted by the variational method 
applied to these set of operators are shown in Fig.~\ref{0+} which shows a 
clear plateau at a value $m(0^+)=0.3629(50)$, in agreement with the naive 
determination of the fundamental mass. Data coming from the next-to-leading
eigenvalue and corresponding to the first excited level are quite noisy. 
Nevertheless, if we take as ``plateau'' value the first value of the effective mass
which agrees at 1$\sigma$ level with the preceding and the following value, we get
$m^*(0^+)=0.684(48)$ and a mass ratio $m^*(0^+)/m(0^+)=1.89(16)$ in good agreement
with the prediction from universality. Further analysis has shown that 
this excited state is coupled with the operators with larger scale in the set
we considered. Indeed, taking into account only two operators (smoothing level 0 
and 15) the same result is obtained. Instead, performing the analysis taking 
only the first four operators (smoothing levels 0, 3, 6, 9) gives $m(0^+)=
0.3615(40)$ and $m^*(0^+)=0.848(28)$, with a mass ratio $m^*(0^+)/m(0^+)=2.35(10)$ 
which is compatible with the value of $m_3/m_1$ in the 3D Ising model and could be 
the analog of the two-particle cut in the same model. 

Finally, in order to look for states with non-zero momentum, we considered
the following set of $2^+$ operators, inspired by Ref.~\cite{ACCH97}, 
\begin{eqnarray*}
\overline P^{(n)}(z)&=&\!\frac{1}{N_x N_y}\sum_{x,y} \biggl[P(x,y,z) P(x+na,y,z)\\
&-& P(x,y,z) P(x,y+na,z)\biggr]\;,
\end{eqnarray*}
with $n=1,\ldots,5$. The variational method gave for the fundamental
$2^+$ mass the value $m(2^+)=1.23(14)$. The ratio of this value to
the fundamental mass in the $0^+$ channel (from the naive determination) 
$m(0^+)=0.3660(20)$ gives $m(2^+)/m(0^+)=3.36(40)$. This result can be
compared with the corresponding mass ratio in the glueball spectrum of the
3D gauge Ising model which, by duality, is connected to the spectrum 
of massive excitations in the broken phase of the 3D (spin) Ising model.
In the 3D gauge model we have $m(2^+)/m(0^+)=2.59(4)$ and $m^*(2^+)/m(0^+)=
3.23(7)$, so our result for the lowest screening mass in the $2^+$ channel
seems to be in agreement with the first excited mass $m^*(2^+)$, more than with
the fundamental $m(2^+)$.


\begin{thebibliography}{9}

\bibitem{CHP99}
M.~Caselle, M.~Hasenbusch and P.~Provero,
Nucl. Phys. B {\bf 556} (1999) 575.

\bibitem{CHPZ01}
M.~Caselle, M.~Hasenbusch, P.~Provero and K.~Zarembo,
Nucl. Phys. B {\bf 623} (2002) 474.

\bibitem{ACCH97}
V.~Agostini, G.~Carlino, M.~Caselle and M.~Hasenbusch, 
Nucl. Phys. B {\bf 484} (1997) 331.

\bibitem{SY82}
B.~Svetitsky and L.G.~Yaffe, Nucl. Phys. {\bf B210} (1982) 423.

\bibitem{MW87}
I.~Montvay and P.~Weisz, Nucl. Phys. B {\bf 290} (1987) 327.

\bibitem{PV91}
R.~Petronzio and E.~Vicari, Phys. Lett. B {\bf 254} (1991) 444.

\bibitem{KP85}
A.D.~Kennedy and B.J.~Pendleton, Phys. Lett. B {\bf 156} (1985) 393.

\bibitem{ES98}
J.~Engels and T.~Scheideler, Nucl. Phys. B {\bf 539} (1999) 557.

\bibitem{Kro90}
A.~S.~Kronfeld, Nucl. Phys. Proc. Suppl. {\bf 17} (1990) 313.

\bibitem{LW90}
M.~L\"uscher and U.~Wolff, Nucl. Phys. B {\bf 339} (1990) 222.

\end{thebibliography}
\end{document}